\documentclass[acmsmall,nonacm]{acmart}
\AtBeginDocument{%
  }

\usepackage{enumitem}
\setlist[itemize]{noitemsep, topsep=0pt}
\usepackage{float}
\usepackage{hyperref}

\renewcommand\footnotetextcopyrightpermission[1]{}

\begin{document}

\title[Fairness and Efficiency in Human-Agent Teams]{Fairness and Efficiency in Human-Agent Teams: An Iterative Algorithm Design Approach}

\author{Mai Lee Chang}
\affiliation{
  \institution{University of Texas at Austin}
  \city{Austin}
  \country{USA}}
\email{mlchang@utexas.edu}

\author{Kim Baraka}
\affiliation{
  \institution{Vrije Universiteit}
  \city{Amsterdam}
  \country{Netherlands}}
\email{k.baraka@vu.nl}

\author{Greg Trafton}
\affiliation{
  \institution{Naval Research Laboratory}
  \city{Washington, DC}
  \country{USA}}
\email{greg.trafton@nrl.navy.mil}

\author{Zach Lalu Vazhekatt}
\affiliation{
  \institution{University of Texas at Austin}
  \city{Austin}
  \country{USA}}
\email{zlv84@utexas.edu}

\author{Andrea Lockerd Thomaz}
\affiliation{
  \institution{University of Texas at Austin}
  \city{Austin}
  \country{USA}}
\email{athomaz@ece.utexas.edu}

\renewcommand{\shortauthors}{Mai Lee Chang et al.}

\begin{abstract}
When agents interact with people as part of a team, fairness becomes an important factor. Prior work has proposed fairness metrics based on teammates' capabilities for task allocation within human-agent teams. However, most metrics only consider teammate capabilities from a third-person point of view (POV). In this work, we extend these metrics to include task preferences and consider a first-person POV. We leverage an iterative design method consisting of simulation data and human data to design a task allocation algorithm that balances task efficiency and fairness based on both capabilities and preferences. We first show that these metrics may not align with people's perceived fairness from a first-person POV. In light of this result, we propose a new fairness metric, fair-equity, and the Fair-Efficient Algorithm (FEA). Our findings suggest that an agent teammate who balances efficiency and fairness based on equity will be perceived to be fairer and preferred by human teammates in various human-agent team types. We suggest that the perception of fairness may also depend on a person's POV.
\end{abstract}



%\received{January 2024}
%\received[revised]{July 2024}
%\received[accepted]{October 2024}

\maketitle

\section{Introduction}

An important component of successful teamwork is the appropriate allocation of tasks among team members. Human-agent teamwork has a history of optimizing for the team's task performance and lack consideration for fairness in task allocations ~\cite{gombolay2017computational, riedelbauch2023benchmarking, natarajan2023human}. Recently, research has proposed fairness metrics for human-agent teaming and investigated the impacts of fairness on the team's performance and relationship dynamics~\cite{jung2020robot, chang2020fairness, chang2021unfair, claure2020multi, habibian2022encouraging, hri2022workshop_fairness, claure2023social}. Along a similar vein, fairness in machine learning (ML) is a rich literature that focuses on designing and evaluating ML systems to reduce bias and discrimination against people~\cite{caton2020fairness, dwork2012fairness, hardt2016equality, hutchinson201950, oneto2020fairness, pessach2022review, pagano2023bias, wan2023processing}. Application areas include healthcare, criminal justice, education, finances, and more~\cite{chen2023algorithmic, le2022survey, deng2022exploring}. 

Can we use ML fairness metrics in human-agent teaming? In human-agent teaming, the main goal is completing the team’s tasks. This is impacted by team members’ contribution which are influenced by factors such as members' capabilities and task preferences. This requires the development of new fairness metrics and leveraging fair ML metrics to human-agent teamwork. In addition, the fair ML literature shows that there tends to be a trade-off between fairness and accuracy (e.g., classification)~\cite{dutta2020there, kim2020fact, menon2018cost, zhao2019inherent}. However, it is uncertain if a similar relationship exist in the human-agent teaming domain for task performance and fairness due to the context sensitive nature of fairness~\cite{runciman1966relative, bicchieri1999local_fairness}. 

In our work, we are interested in task allocation in a team composed of an agent and multiple humans. Different from prior work, we are interested in the influences of teammates' traits, including their capabilities and preferences and their perception of fairness. This is a complex problem that involves an understanding of the relationships between capabilities and preferences at both the individual and team level as well as the perspective of all stakeholders who may be judging fairness. 

A small body of work in human-agent teaming have explored notions of fairness based on workload, teammate capabilities, task type, and the amount of time the agent spends working with human teammates in the setting of task allocations~\cite{chang2020fairness, chang2021unfair, claure2020multi}. Prior work also investigated the impact of  unequal distribution of tasks on the team's interpersonal relationships~\cite{jung2020robot, claure2023social} and trust in the agent~\cite{claure2020multi}. 

Previous research suggests that people's point of view (POV), e.g., first-person, third-person, influences the impact of fairness in task allocations in human-agent teams~\cite{habibian2022encouraging}. Existing research showed that fairness based on capabilities aligns with human teammates' perception of fairness from a third-person POV such as a supervisor~\cite{chang2021unfair}. We advance on this by exploring if this finding holds from a first-person POV.

%Prior work has developed fairness metrics and algorithms for fair allocation of resources (e.g., room, rent) and indivisible goods. These fairness metrics include envy-free, ect... 

Our research questions include:
\begin{enumerate}
    \item Can we extend prior fairness metrics that has been shown to align with people's perception of fairness from a first-person POV to a third-person POV?
    \item Can agents balance task efficiency and fairness in human-agent teamwork? %(i.e., show the benefits of including fairness in the algorithm  fairness is not just nice to have) 
\end{enumerate}

We use an iterative algorithm design approach that is inspired by the core practice of human-computer interaction (HCI) design of rapid prototyping and testing~\cite{yang2020re}. Our approach favors small sample size user testing in combination with simulation studies, in contrast with a large sample size experimental approach. We believe the emphasis on the process of developing an algorithm for fairness in human-agent teams generates insights that are equally important as the final outcome, and is of value to the human-agent interaction (HAI) community.

Using this iterative design approach, we design the Fair algorithm that enables the agent to consider fairness based on teammates' capabilities and task preferences. We evaluate the Fair algorithm against the baseline algorithm that optimizes for solely task efficiency in a simulation study and then in a user study with 6 participants. We discovered limitations of the existing metric that is based on capabilities. We leverage the fairness concept of fair-equity from sociology~\cite{adams1965inequity} and fair ML~\cite{corbett2017algorithmic} to develop the Fair-Efficient Algorithm (FEA) that incorporates a new fairness metric based on equity and evaluate it in both simulation and with participants ($n=7$). Our results suggest that FEA is able to balance task efficiency and fairness. 

\section{Related Work}

\subsection{Fairness in Human-Agent Teams}\label{sec:lit_fair_hrt}
Recent research started exploring fairness in the context of task allocations within human-agent teams~\cite{jung2020robot, chang2020fairness, chang2021unfair, claure2020multi, habibian2022encouraging, hri2022workshop_fairness, claure2023social}. For example, researchers have shown that an agent's effortful and fluent behavior increases the human teammate's perception of fairness in the team~\cite{chang2020fairness}. Previous work in human-agent teaming proposed fairness metrics based on workload, teammate capabilities, task type, and the amount of time the agent spends working with human teammates~\cite{chang2020fairness, chang2021unfair}. Prior studies showed that an unequal distribution of tasks among team members negatively impacted relationship satisfaction in human-agent teamwork~\cite{jung2020robot, claure2020multi}. Claure et al.~\cite{claure2023social} found that the type of allocator (human vs. agent) did not significantly impact teammates' perception of fairness. However, these approaches do not consider granular skill levels and fairness based on task preferences since the literature shows that task preferences is important for safe, effective human-agent teaming~\cite{nikolaidis2013crosstraining, gombolay2017computational}. 

\textbf{Point of View:} Moreover, prior research suggests that people's point of view (POV) influences the importance of fairness in task allocations in human-agent teams~\cite{habibian2022encouraging}. An example of a first-person POV is a team member's assessment of the task allocation whereas a third-person POV is a supervisor's assessment. Existing research suggests that fairness matters more for first-person POV than for third-person POV~\cite{habibian2022encouraging}. Chang et al.~\cite{chang2021unfair} assessed fairness based on teammates' capabilities from a third-person POV in a single-agent multiple-human team. In this work, we are interested in assessing this type of fairness from a first-person POV. 

\textbf{Trade-offs With Fairness:} The fair ML literature shows that there is typically a trade-off between fairness and accuracy~\cite{dutta2020there, kim2020fact, menon2018cost,zhao2019inherent}. However, it is uncertain if a similar relationship exist in the human-agent teaming domain for task performance and fairness due to the context sensitive nature of fairness~\cite{runciman1966relative, bicchieri1999local_fairness}. In this paper, we explore if there is a trade-off between task efficiency and fairness.

\subsection{Human-Agent Team Types}
Human-agent teams can be defined by composition and size (number of agents, number of people) and the agent's role~\cite{sebo2020robots_team_review}. The human-agent team composition and size that this paper cover is single agent - groups of humans (two). Human-agent teaming research to date has focused mainly on single agent - single human teams~\cite{sebo2020robots_team_review, wolf2020human}. When an agent interacts with groups/teams of people, this introduces more challenges for the agent. For instance, the agent's actions influence both individual team members and the overall team dynamics and outcomes~\cite{jung2020robot}. The interactions in the team becomes more complex because besides HAI, there exist also human-human interactions~\cite{sebo2020robots_team_review}.

Agents can take on the role of a follower, peer, or leader~\cite{sebo2020robots_team_review}. A follower reacts to interaction initiatives from people or follows people's instructions. A peer has the same level of decision-making authority as people. A leader initiates or guides interactions in the team. In this paper, the agent is a leader who allocates tasks to the human teammates. We advance the literature by understanding how to balance fairness and task efficiency within different human-agent team types. 

%%%%%%%%%%%%%%%%%%%%%%%%%%%%%%%%%%%%%%%%%%%%%%%%%%%%%%%%%%%%%%%%%%%%%%%%
\section{Problem Formulation}

\begin{figure}[!thb]
\centering
\includegraphics[width=0.8\linewidth]{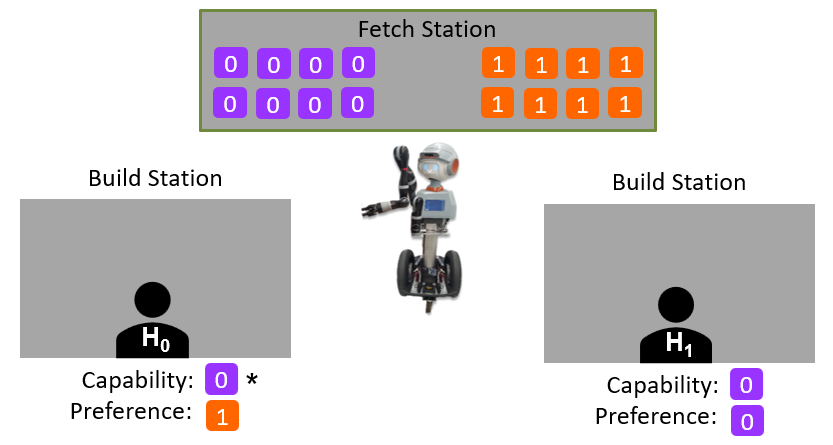}
\caption{An example task allocation scenario that falls under our problem formulation. There are different task types (e.g., 0 and 1) and the human teammates may differ in their capabilities and task preferences.}
\label{fig:cookie_scenario}
\end{figure}

We consider a task allocation where there is one robot (allocating tasks) and two humans (performing tasks; adapted from \cite{gombolay2015preference}) as shown in Figure \ref{fig:cookie_scenario}, and we consider a finite horizon. In most team settings, including this one, fairness matters because people have preferences (what type of task they prefer) and capabilities (how efficient they are). 

At each time step or round, the robot delivers kits to both humans (each human receives a kit). A kit belongs to a task type (e.g., squares cookies, letters cookies). Each time step is a fixed duration so people continuously work until time is up. Each kit has an infinite number of tasks such as an infinite number of squares cookies. At the next time step, the robot delivers kits again and this repeats until all the kits are allocated. The kit that the robot allocates to the humans is based on the optimal policy. 

We model the interaction as a Markov Decision Process (MDP), which is defined as a tuple $M = \langle S, A, T, R, \gamma \rangle$. $S$ is a finite set of states, with possible goal state $G \subset S$. $S$ is described by the number of kits that are in the fetch station ($T_f$) , number of kits that are in human 0's station ($T_{H_0}$) and number of kits that are in human 1's station ($T_{H_1}$) for all the task types. $T_f$ means these are the remaining kits that still need to be allocated. $T_{H_0}$ means these are the kits that are allocated to human $0$ and similarly for $T_{H_1}$. $G$ is any state where $T_f = 0$. $A_R$ is the set of robot actions. A robot action $a_R \in A_R$ is fetching human 0 a task type ($a_0$) and fetching human 1 a task type ($a_1$) for all task types.

\begin{itemize}
\item $S = [T_f, \; T_{H_0}, \; T_{H_1}] \; \forall \; \text{task types}$.
\item $G = [0, \; T_{H_0}, \; T_{H_1}] \; \forall \; \text{task types}$.
\item $A_R \in \{a_0, \; a_1\} \; \forall \; \text{task types}$.
\end{itemize}

\noindent
The transition function is denoted by $T$. The reward function $R$ depends on the algorithm's objective as defined in Section~\ref{sec:sim1_reward_signal}. The discount factor $\gamma$ is in the range $[0, 1]$.

\subsection{Fairness Definitions}\label{sec:ec_ep_definitions}

We define fairness ($F_f$) based on people's features, $f$. In this paper, we focus on the features of capability ($c$) and task preferences ($p$). We expand the fairness definition of equality of capability ($F_c$)~\cite{chang2021unfair} to include granular skill levels since it only considers the task type that each teammate is the best at. $F_c$ equalizes team members' capabilities among them. In a similar manner, we create equality of preference, $F_p$, to account for granular levels of people's task preferences. We define $F_f$ of state $s$ as follows: 
\begin{equation}
    F_f(s) = \frac{\sum\limits_{j=1}^n (f_{H_i, j} * T_{H_i, j}(s) - f_{H_{i+1}, j} * T_{H_{i+1}, j}(s))}{\frac{1}{h}\sum\limits_{j=1}^n (T_{H_i, j}(s) + T_{H_{i+1}, j}(s))},
\end{equation}
\noindent where $f_{H_i,j}$ is human $i$'s ($H_i$) feature (e.g., capability, preference) on task type $j$. $f_{H_i,j}$ is in the range [0, 1] where higher is better. $T_{H_i, j}(s)$ is the total number of kits allocated to $H_{i}$ that are of task type $j$ in state $s$. The total number of human teammates is denoted by $h$. $F_f(s)$ is in the range [-1, 1].

The fair metric $F$ for state $s$ is defined as:
\begin{equation}\label{eq:F_s}
    F(s) = \bigg |\frac{F_c(s) + F_p(s)}{h}\bigg |
\end{equation}
In Equation~\ref{eq:F_s}, we use absolute value because we are interested in when any unfairness occurs. It does not matter who the unfairness is towards. Using the absolute value for fairness is an approach that is also used in fair ML such as disparate impact~\cite{corbett2017algorithmic}. Next, we incorporate these fairness definitions in the reward function.

%%%%%%%%%%%%
\subsection{Reward Function Design}\label{sec:sim1_reward_signal}

We design two reward functions. The baseline reward function optimizes for efficiency and the other reward function optimizes for fairness. First, we define the efficiency of state $s$, $E(s)$, as the total contribution of the team based on capabilities. $E(s)$ is normalized by the total number of tasks completed by the team and is in the range 0 to 1. $E(s)$ is defined as:
\begin{equation}\label{eq:efficiency}
    E(s) = \frac{\sum\limits_{j=1}^n (c_{H_i, j} * T_{H_i, j}(s) + c_{H_{i+1}, j} * T_{H_{i+1}, j}(s))}{\sum\limits_{j=1}^n (T_{H_i, j}(s) + T_{H_{i+1}, j}(s))}.
\end{equation}

\noindent
The reward function for each algorithm is in Table~\ref{tab:reward_fcts_1}. An invalid action results in staying in state $s$, i.e., probability of transition is equal to 0. We seek to understand how well the Fair algorithm captures people's perceived fairness from a first-person POV in comparison to the Efficient algorithm. 

\begin{table}[!htb]
\centering
\begin{tabular}{|p{6cm}|p{6cm}|}
 \hline
  Efficient algorithm: $R(s)=$ & Fair algorithm: $R(s)=$ \\ 
%\hhline{|=|=|}%\hline\hline
\begin{minipage}{6cm}
\begin{equation}\label{eq:eff_reward_signal}
\begin{cases}
E(s) & s \in S_{goal}, \\
-1 & \textit{a is invalid in s}, \\
0 & otherwise.
\end{cases}\\
\end{equation}
\end{minipage} & 
\begin{minipage}{6cm}
\begin{equation}\label{eq:fair_ec_ep_reward}
\begin{cases}
1-F(s) & s \in S_{goal}, \\
-1 & \textit{a is invalid in s}, \\
0 & otherwise.
\end{cases}\\
\end{equation}
\end{minipage} \\
\hline
\end{tabular}
\caption{Reward function for each algorithm.} \label{tab:reward_fcts_1}
\end{table}

\begin{comment}
\textbf{Efficient algorithm} is defined as:
\begin{equation}\label{eq:eff_reward_signal}
R(s) = \begin{cases}
E(s) & s \in S_{goal}, \\
-1 & \textit{a is invalid in s}, \\
0 & otherwise.
\end{cases}
\end{equation}

\noindent
The reward function for the \textbf{Fair algorithm} is defined as:
\begin{equation}\label{eq:fair_ec_ep_reward}
R(s) = \begin{cases}
1-F(s) & s \in S_{goal}, \\
-1 & \textit{a is invalid in s}, \\
0 & otherwise.
\end{cases}
\end{equation}
\noindent
\end{comment}

\section{Methodology Overview}
We use a rapid design and testing approach to understand how to formalize fairness in the reward function. We perform a set of simulation studies that are complemented by small sample user testing.

%%%%%%%%%%%
\subsection{Simulation Study Design Overview}\label{sec:sim_studies_overview}
 
There are some situations and team types where fairness may not be a critical factor to team success (e.g., if teammates have balanced capabilities and different preferences).
Thus, our goals of the simulation study are 1) understand when fairness is likely to matter 
and 2) use simulated data to create the agent teammate to focus on certain team types in the user studies (Section~\ref{sec:design_agent}). We simulated 500,000 teams by randomly drawing each team members' capability coefficients and preference coefficients from a uniform distribution within the interval [0.01, 0.99]. For each team, we use policy iteration to solve for the optimal policies. Each team has one policy from the Efficient reward function and another policy from the Fair reward function as defined in Section~\ref{sec:sim1_reward_signal}.

Our evaluation metrics are the fair gap and efficiency gap. For each team, the fair gap is the difference in fair rewards which is equal to the fair reward (Equation~\ref{eq:fair_ec_ep_reward}) of the Fair algorithm minus the fair reward of the Efficient algorithm. A positive fair gap means that the Fair algorithm is fairer than the Efficient algorithm. The efficiency gap is equal to the efficient reward (Equation~\ref{eq:eff_reward_signal}) of the Fair algorithm minus the efficient reward of the Efficient algorithm. A positive efficiency gap means that the Fair algorithm is more efficient than the Efficient algorithm.  

For implementation of the algorithms, the settings for our MDP solver, MDPtoolbox \cite{chades2014mdptoolbox}, are the same for all evaluations: the discount factor is set to 0.9, the stopping criterion is set to 0.0001, and the maximum iterations is set to 100. We solve the MDP by using policy iteration. The transition and reward functions are deterministic for all tasks in this work.

%%%%%%%%%%%%
\subsection{User Studies Design Overview}\label{sec:user_studies_overview}

We provide a brief description of the design of the user studies here. Our studies were approved by an Institutional Review Board. 

\subsubsection{Participants}
\begin{figure}[!thb]
\centering
\includegraphics[width=0.8\linewidth]{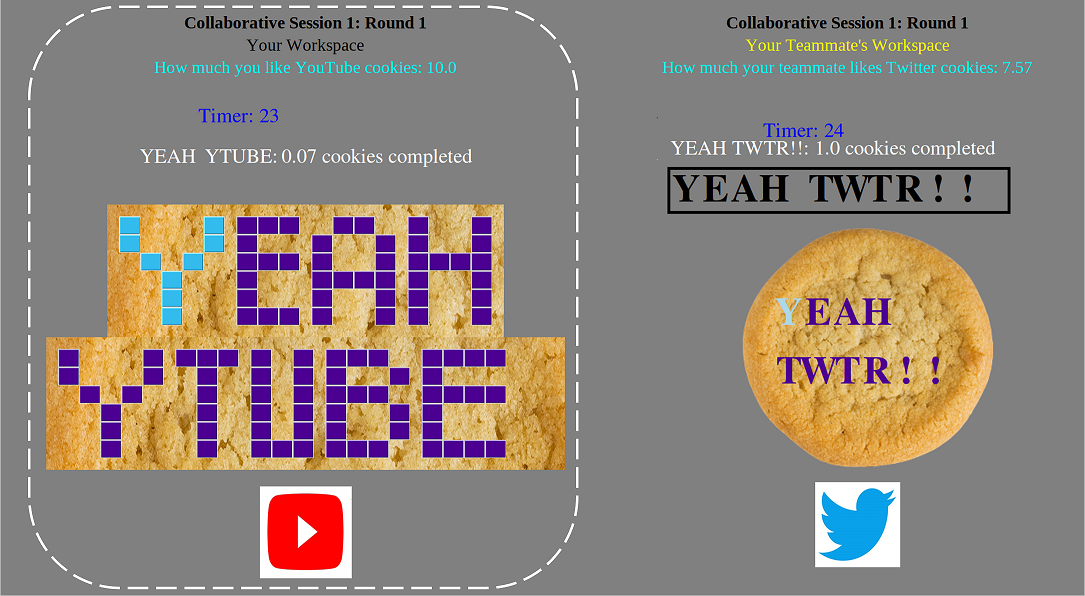}
\caption{User Study \#1: An example of the collaborative session showing each teammate's workspace and the two task types used in the user studies: decorate squares cookies (left) and decorate letters cookies (right).}
\label{fig:collab_screen_example}
\end{figure}

Participants worked in a virtual bakery performing cookie decorations. There were two types of tasks: letter cookies and square cookies; see Figure~\ref{fig:collab_screen_example}. The letter cookies were easy to decorate (participants could drag entire letters to form words). The square cookies were more difficult to decorate (participants needed to click individual boxes to form letters). Cookies were decorated with specific themes (e.g., ``Go Ravens!'').  The robot allocated cookie types to each participant; their goal was to complete as many cookies as possible within an alloted time. Note that the robot changed allocations based on the condition.

\textbf{Capabilities:}
Participants' capabilities were assessed before the experimental trials began. Capability coeficients for both square and letter cookies were collected by calculating their average speed to decorate each cookie type.

\textbf{Preferences:}
In addition, participants' preferences were assessed before the experimental trials began. Preferences were collected by presenting themes that participants were likely to feel strongly about (e.g., a local football team and their closest rival) with the goal of finding a range of preferences.

Cookies could thus be created to have a positive (high capability, high preference) or negative (high capability, low preference) relationship.

\subsubsection{Design of Agent Teammate}\label{sec:design_agent}
Since we know a participant's capabilities and preferences before the experimental trials began, we can design an agent teammate with specific characteristics to fit a specific team type (e.g., teams that meet a certain fair gap criteria, specific relationships between team members' capabilities and preferences). After the practice session, the participants' capability coefficients and preference coefficients are used to select the agent teammate's coefficients from the simulated teams. This method reduces the computation time to find the teammate's coefficients. From a specified set of simulated teams, we find a team where one of the members' coefficients matches the participants' coefficients the best. We minimize the distance ($L_1$) between participant $i$'s coefficients ($\mathbf{y_i}$) and the simulated team member $i$'s coefficients ($\hat{\mathbf{y_i}}$) as defined in Equation~\ref{eq:l1}:
\begin{equation}\label{eq:l1}
    L_1 = \sum\limits_{i=1}^n |\mathbf{y_i} - \hat{\mathbf{y_i}}|.
\end{equation}
For the simulated team member that is closest to the participant (smallest $L_1$ value), we use this simulated team member's teammates' coefficients for the agent teammate. 

\subsubsection{Study Design}

This was a within experiment with two conditions. In Study \#1, the conditions were the Efficient algorithm and Fair algorithm. In Study \#2, the conditions were the Efficient algorithm and FEA. Participants performed 8 rounds of each condition. When participants arrived, they were told they would be collaborating with a simulated robot and another participant, who they met. In actuality, the agent teammate described above was the actual participant's teammate, though participants believed it was the human confederate. Participants were debriefed and told that their teammate was an agent. 

The independent variable is the algorithm type (e.g., Fair vs. Efficient). The dependent variables include both objective and subjective measures. The objective metrics are total cookies completed by the team and specific types of rounds allocated to each teammate. The subjective metrics are participants' agreement with the fairness metrics and their assessment of the teamwork based on the fairness metrics. All measures are described in Table~\ref{tab:pilot1_questionnaire1}.

\begin{table}[!htb]
\centering
\begin{tabular}{p{5in}}
\hline
%\textbf{Instructions:} \\
%Please answer the questions below based on this collabortive session.
%In this questionnaire, "best at" is referred to as the decoration cookie speed.
%"Prefer" is how much you like each themed cookie (i.e., Dogs, Snakes).\\
%\hline\hline
\textbf{Fairness based on capabilities:} \\
1. [Cap-Agree] In general, working on tasks that you are best at is a fair way to contribute to the team. (Likert)\\
2. [Cap-Rating] The robot was fair to both my teammate and I with respect to us working on tasks that we are best at. (Likert)\\
\hline
\textbf{Fairness based on preferences:} \\
3. [Pref-Agree] In general, working on tasks that you prefer is a fair way to contribute to the team. (Likert)\\
4. [Pref-Rating] The robot was fair to both my teammate and I with respect to us working on tasks that we prefer. (Likert)\\
\hline
\textbf{Overall fairness:} \\
5. [Fair-Overall] Overall, the robot allocated the tasks fairly between my teammate and me. (Likert)\\
6. Please elaborate on your answer to question \#5 above.\\
\hline
\end{tabular}
\caption{Questionnaire administered after each collaborative session (7-point Likert items).
\label{tab:pilot1_questionnaire1}}
\end{table}

\subsubsection{Study Procedure}\label{sec:user_study_procedure}
The general structure of the user study is as follows:
\begin{enumerate}
    \item Consent form,
    \item Information about cookie decoration themes,
    \item Survey about participant's preferences,
    \item Practice session,
    \item Collaborative Session 1 (Efficient algorithm): 8 rounds and survey,
    \item Collaborative Session 2 (Fair algorithm or Fair-Efficient algorithm): 8 rounds and survey,
    \item Post-study survey,
    \item Debrief.
\end{enumerate}

When the participant arrives, they are informed that they will be collaborating with a simulated robot and another participant. This other participant is a confederate who is only acting to be the participant's teammate. In reality, participants collaborate with an agent teammate. The rationale for designing the study this way is that we are interested in studying a specific set of teams. The confederate is needed for participants to view their teammate as a person instead of an agent. 

After obtaining informed consent, the experimenter then explains that there will be a lot of mouse clicking so to minimize distractions, the other participant/confederate's work station is in the room next door but the programs will sync and they will be able to see each other working during the collaborative sessions.

Next is the survey asking participants to rate their preference of cookie decoration themes. The practice session is composed of four rounds, each round lasting 30 seconds. Two rounds are for practicing making the squares cookie and the other two rounds are for the letters cookie. Data from the second rounds are used to calculate the person's capability coefficients which are used in the algorithms. 

Following this, the first round of Collaborative Session 1 begins. Each session consists of eight rounds where each round lasts up to 30 seconds. The sessions differ in which simulated robot/algorithm was used. During the collaborative sessions, the participant is able to see their teammate working (Figure~\ref{fig:collab_screen_example}). 

Following each of these collaborative sessions, the participant fills out a questionnaire (Table~\ref{tab:pilot1_questionnaire1}) asking them to rate their level of agreement with each of the fairness metrics: (1) fairness based on capabilities, (2) fairness based on preferences, and (3) overall fairness. We also asked participants to assess the robots based on the fairness metrics. The study concludes with a debrief session in which the experimenter reveals that the participant's teammate was a agent and obtains consent to use the data.

%%%%%%%%%%%%%%%%%%%%%%%%%%%%%%%%%%%%%%%%%%%%%%%%%%%%%%%%%%%%%%%%%%%%%%%%
\section{Evaluation of Fair Algorithm}\label{sec:results_1}

\subsection{Simulation Study \#1}\label{sec:sim1_results_fair}

Since we are interested in teams where the Fair Algorithm outperforms the Efficient algorithm in terms of fair rewards, we filter for teams with a certain fair gap. We are interested in a fair gap greater than 0.60. We hypothesize that with this fair gap, participants will perceive the Fair algorithm to be fairer than the Efficient algorithm. After filtering, 7,072 teams remained. Table~\ref{tab:sim1_results_fair} show a summary of the distribution of the filtered teams with respect to the fair gap. About $75\%$ of the filtered teams have a fair gap that is within the interval [0.60, 0.70).

\begin{table}[t]
\centering
\begin{tabular}{ p{.12\linewidth}|c|c|c|c|c|c|c|c|c|c|c|c}
 \hline
 \multicolumn{1}{c|}{} & \multicolumn{3}{|c|}{Fair Gap}\\
 \hline
   & [0.60, 0.70) & [0.70, 0.80) & [0.80, 0.90] \\
  \hline
  Teams & 
  5,302 (74.97\%) & 1,566 (22.14\%) & 204 (2.80\%)\\
\hline
\end{tabular}
\caption{Simulation study \#1 results: Count of teams with a fair gap greater than 0.60. \label{tab:sim1_results_fair}}
\end{table}

For the set of filtered teams with a fair gap greater than 0.60, about $40\%$ of the filtered teams have an efficient gap that is within the interval [-0.10, 0.00] meaning that the Efficient algorithm is the same or better than the Fair algorithm in terms of efficient rewards. The rest of the filtered teams have an efficient gap that is within the interval [-0.50, -0.10). Also, from this set of filtered teams, we design the agent teammate's behavior in the user study as described in Section~\ref{sec:design_agent}.

%The top plot in Figure~\ref{fig:sim1_results} shows the distribution of the set of teams (total of 7,072 teams) that have a fair gap greater than 0.60. 

% fair gap frequency: 5302, 1566,  199,    5]
% efficient gap frequency:  3,   32,  229,  595, 1300, 2076, 2837

% how to read bins in figure: first bin is [0.60, 0.70) (including 0.60, but excluding 0.70) and the second [0.70, 0.80). The last bin, however, is [0.80, 0.90], which includes 0.90.

%%%%%%%%%%%%
\subsection{User Study \#1}\label{sec:pilot_user1}
A total of six participants (1 female, 5 males, age: $M = 28.00$), participated in the study, which lasted about 30 minutes. Due to the small sample size, we do not perform statistical tests but plot the data to see the trends. From the practice session results, participants' average capability coefficient for the squares cookie is $0.26$. Participants' average capability coefficient for the letters cookie is $0.92$. Participants' average preference coefficient for the squares cookie is $0.42$ and for the letters cookie is $0.54$. Note that higher capability coefficients mean more capable and higher preference coefficients mean more preferred. In regards to how well we were able to select the agent teammate, the average $L_1$ value is $0.20$ where closer to 0 means better selection.

\subsubsection{Task Performance}
Teams interacting with the Efficient algorithm seem to complete more cookies on average ($24.55$) than teams interacting with the Fair algorithm ($19.23$).

\subsubsection{Perceived Fairness}

\begin{figure}[!thb]
\centering
\includegraphics[width=0.8\linewidth]{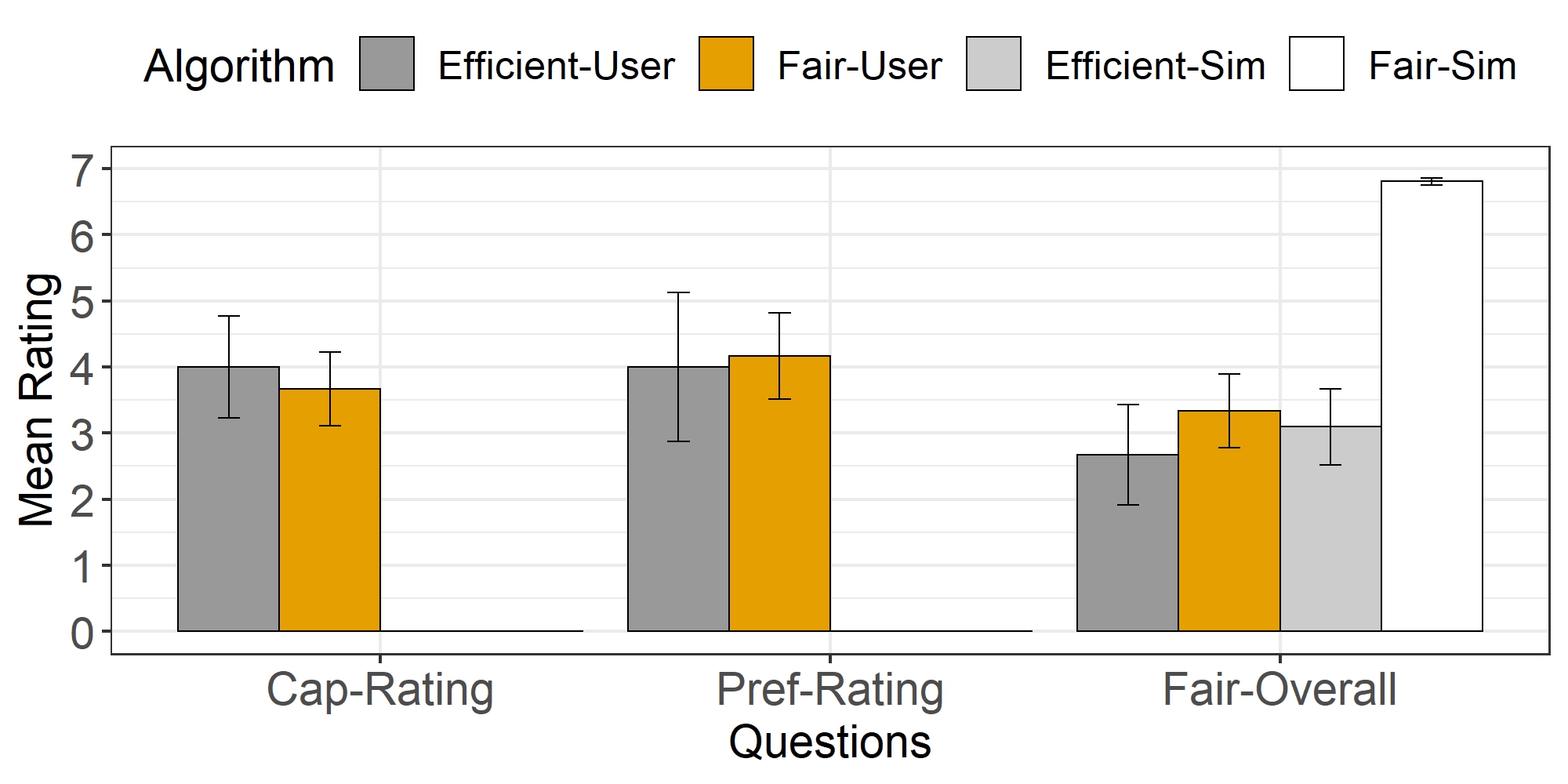}
\caption{Predicted fairness (scaled to rating between 1 and 7) from Simulation Study \#1 and participants' ratings from User Study \#1: Higher ratings mean more fair. Please see Table\ref{tab:pilot1_questionnaire1} for the specific questions. Cap is the abbreviation of capability. Pref is the abbreviation of preference.}
\label{fig:user1_ratings}
\end{figure}

Figure~\ref{fig:user1_ratings} shows participants' average fairness ratings and the predicted fairness from Simulation Study \#1. The predicted fairness is the fair reward scaled to rating between 1 and 7.

\textbf{Fairness Based on Capabilities:}
Participants interacting with both algorithms seem to highly agree that in general, working on tasks that you are best at is a fair way to contribute to the team (Efficient: $M = 5.00$; Fair: $M = 6.00$). When asked about their rating of equality of capability, participants seem to feel that the Efficient algorithm was fairer (Efficient: $M = 4.00$; Fair: $M = 3.67$); Figure~\ref{fig:user1_ratings}.

\textbf{Fairness Based on Preferences:}
In terms of agreement about fairness based on preferences as a metric, participants in both conditions seem to agree (Efficient: $M = 4.83$; Fair: $M = 4.33$). Participants seem to perceive both algorithms to be similar (Efficient: $M = 4.00$; Fair: $M = 4.17$) in regards to fairness based on preferences, which contradicts our expectations (Figure~\ref{fig:user1_ratings}).

\textbf{Overall Perceived Fairness:}
For overall perceived fairness, participants seem to rate both algorithms on the lower end (Efficient: $M = 2.67$; Fair: $M = 3.33$), which did not align with our expectations since the predicted fairness rating from our simulation study is $7$.

\subsection{Summary: Lessons Learned}\label{sec:pilot1_lessons_learned}
Taking into account the insights gained from the first set of simulation and user studies, we present the lessons learned about the design of the algorithms and metrics.

\textbf{1. Equality of Capability:} Our insights suggest a reconsideration of the $F_c$ metric. Participants who are the most capable in the team want to be given more opportunities to work on the tasks that they are best at which is a contrast to Chang et al.'s findings~\cite{chang2021unfair}. Their findings were based on an assessment of $F_c$ from a third-person POV whereas in this paper, $F_c$ is assessed from a first-person POV. This difference in results aligns with prior work suggesting that the POV impacts perceived fairness~\cite{habibian2022encouraging}. Currently, $F_c$ reduces the opportunities for the most capable team member because the metric is trying to balance the capabilities contributions across the team mates. This led us to look into the social science literature about other allocations concepts (Section~\ref{sec:implement_lesson_fair_equity}).

\textbf{2. Relationship Between Efficiency and Fairness:} Our insights seem to indicate that people prioritize team efficiency but they still value fairness. This led us to review literature about the relationship between efficiency and task preferences in teamwork. We implement this lesson learned in Section~\ref{sec:implement_lesson_reward}.

\textbf{3. Team Types:}
Our results highlight the need to understand team members' perception of fairness in different team types. The characteristics of team types include the relationship between a team member's capabilities and preferences and which team member is the most capable in the team. These characteristics may impact perceived fairness. We define the various team types in Section~\ref{sec:team_types}.

\textbf{4. Additional Objective Metrics:}
The total cookies completed by the team is a metric of the team's task performance but additional objective metrics are needed to gain a better understanding of the allocations in terms of capabilities and preferences. These additional metrics would need to be different from the rewards. Section~\ref{sec:metrics_allocations} covers these metrics.

\textbf{5. Transparency of the Algorithm:} Participants' open-ended comments reveal that they assess how well the Fair algorithm does in terms of fairness based on other factors besides capabilities and preferences. The main factor mentioned is task difficulty. This finding aligns with prior work showing that there are bleedover effects in people's assessment of fairness~\cite{chang2021unfair}. This suggests that people's assessments are unfair for the Fair algorithm since it is only designed to reason about capabilities and preferences and not other factors. One way to calibrate participants' mental models is to be transparent about what each algorithm is considering~\cite{de2007effects}.

%%%%%%%%%%%%%%%%%%%%%%%%%%%%%%%%%%%%%%%%%%%%%%%%%%%%%%%%%%%%%%%%%%%%%%%%
\section{Implementation of Lessons Learned}\label{sec:implement_lessons_learned}

\subsection{Design of Fair-Equity Metric}\label{sec:implement_lesson_fair_equity}
Results from our first set of evaluations (Sections~\ref{sec:results_1}) led to the development of a new fairness metric called fair-equity. Fair-equity is motivated by the Theory of Equity~\cite{adams1965inequity}. Adams' Theory of Equity states that an equitable distribution is where the ratio of one person's outcomes to inputs is equal to the ratio of the other person's outcomes to inputs~\cite{adams1965inequity}. Outcomes represent rewards such as pay, seniority benefits, and job status. Inputs represent the contributions that the person brings to the job such as experience, education, and physical effort. Inequity results when a person is so to speak relatively underpaid (high effort and low pay) and also when a person is relatively overpaid (low effort and high pay). Prior studies provide support for Adams' Theory of Equity~\cite{adams1965inequity, goodman1971examination, lane1971equity}. 

In this paper, outcomes is represented by the preference score. Human $i$'s outcomes is defined as the sum of the product of $H_i$'s preference of task type $j$ ($p_{H_i, j}$) and the total number of tasks completed by $H_i$ that are of task type $j$ ($T_{H_i, j}$):

\begin{equation}\label{eq:outcomes}
    H_{i, outcomes} = \sum\limits_{j=1}^n (p_{H_i, j} * T_{H_i, j}(s))
\end{equation}
\newline
Inputs is presented by the total capabilities for task types $j$ and defined as:
\begin{equation}\label{eq:inputs}
    H_{i, inputs} = \sum\limits_{j=1}^n c_{H_i, j}
\end{equation}
\newline
The fair-equity ($F_{E}$) metric for state $s$ is defined as:
\begin{equation}\label{eq:fair_equity}
    F_E(s) = \frac{H_{i, outcomes}}{H_{i, inputs}} - \frac{H_{{i+1}, outcomes}}{H_{{i+1}, inputs}} 
\end{equation}
\newline
Notations are defined in Section~\ref{sec:ec_ep_definitions}.
%\begin{equation}\label{eq:fair_equity}
    %F_E(s) = \frac{\sum\limits_{j=1}^n (p_{H_i, j} * T_{H_i, j}(s))}{\sum\limits_{j=1}^n c_{H_i, j}} - \frac{\sum\limits_{j=1}^n (p_{H_{i+1}, j} * T_{H_{i+1}, j}(s))}{\sum\limits_{j=1}^n c_{H_{i+1}, j}}
%\end{equation}

\subsection{Design of New Reward Function}\label{sec:implement_lesson_reward}

Implementing our lessons learned regarding a re-evaluation of the equality of capability metric along with examining prior literature about the relationship between task efficiency and preferences, we create a new reward function. Prior work in human-agent teaming shows that people prioritize team efficiency~\cite{gombolay2015preference, gombolay2015decision}. That is, they want robotic teammates to consider their task preferences only if it does not negatively impact the team's efficiency "too much"~\cite{gombolay2015preference,gombolay2015decision}. It's unclear what "too much" means, but we can draw some insights from Claure et al.'s work on a multi-arm bandit algorithm with fairness constraints~\cite{claure2020multi}. When the weak players got to play at most $25\%$ of the time, they perceived the allocation to be unfair whereas the strong players perceived it to the fair. When the weak players got to play at most $33\%$ of the time, both players perceived the allocation to be fair. Note that not getting to play means being idle. Taking into account this exiting work and our fair-equity metric, we design the Fair-Efficient Algorithm (FEA) that considers both task efficiency and fair-equity. The reward function for FEA is defined as: 

\begin{equation}\label{eq:eff_fair_reward_signal}
R(s) = \begin{cases}
\lambda E(s) + (1-\lambda)(1-|F_E(s)|') & s \in S_{goal}, \\
-1 & \textit{a is invalid in s}, \\
0 & otherwise.
\end{cases}
\end{equation}

\begin{equation}
    |F_E(s)|' = \frac{|F_E(s)|-min(|F_E(s)|)}{max(|F_E(s)|) - min(|F_E(s)|)}.
\end{equation}

\noindent
$|F_E(s)|'$ is $|F_E(s)|$ scaled to the range $[0, 1]$ using min-max scaling. The $min$ and $max$ are over the set of user data (Section~\ref{sec:pilot_user1}). We use $|F_E(s)|$ because we are interested in when any unfairness occurs. $\lambda$ is a weight on efficiency and is in the range $[0, 1]$. Note that the efficiency of a state, $E(s)$, is defined as in Equation~\ref{eq:efficiency}. The fair-equity of a state, $F_E(s)$ is defined as in Equation~\ref{eq:fair_equity}. For all studies in this paper, $\lambda$ is set to 0.70.

%%%%%%%%%%%
\subsection{Human-Agent Team Types}\label{sec:team_types}

Based on our lesson learned about the need to understand the impact of team types on perceived fairness, we define human-agent team types based on the relationship between teammates' capabilities and task preferences. A human teammate's capabilities and preferences can be \textit{positively} correlated, that is, they prefer to work on tasks that they are capable. On the other hand, a human teammate's capabilities and task preferences can be \textit{negatively} correlated. For example, a person may want to work on a new task to improve their skills or reduce boredom. In each team type, there is at least one person who is the most capable in the team. We define three team types. In the \textbf{Mixed team type}, one human teammate's capabilities and task preferences are \textit{negatively} correlated and the other human teammate's capabilities and task preferences are \textit{positively} correlated. In the \textbf{Twins team type}, both human teammates' capabilities and preferences are positively correlated. Lastly, in the \textbf{Negative team type}, booth human teammates' capabilities and preferences are negatively correlated. 

In our work, we assume a prior that each human teammate has a 50\% chance of having a positive correlation between their capabilities and preferences and 50\% chance of having a negative correlation. This means that 50\% of the teams belong to the Mixed team type, 25\% of the teams belong to the Twins team type and 25\% of the teams belong to the Negative team type. In this paper, we investigate the Mixed and Twins team types.

%%%%%%%%%
\subsection{Metrics for Allocations}\label{sec:metrics_allocations}
One of the lessons learned is the need to have additional objective metrics to assess fairness of the allocations. Taking this into account, our new objective metrics include the average number of rounds that each team member is allocated tasks that they were most capable at and the average number of rounds that each team member is allocated tasks that they preferred the most.

%%%%%%%%%
\subsection{Transparency of Algorithms}\label{sec:implement_transparency}
To help participants have an accurate model of the design of the algorithms, we modified the instructions to include explicit information about the algorithms: ``Robot Chris considers only the decoration cookie speeds (yours and your teammate's) in its assignments. Robot Pat considers both the decoration cookie speeds and preferences of each themed cookie (yours and your teammate's)''. This modification is based on prior research showing that the transparency of procedural fairness can positively impact cooperation~\cite{de2007effects}. Also, we added a post-study questionnaire asking participants to assess the algorithms only along the dimensions that it was designed to consider: ``Which robot did a better job considering your teammate's and your preferences of each themed cookie and decoration speeds in its assignments?'' (7-point Likert items).

%%%%%%%%%%%%%%%%%%%%%%%%%%%%%%%%%%%%%%%%%%%%%%%%%%%%%%%%%%%%%%%%%%%%%%%%
\section{Evaluation of FEA}\label{sec:implement_2}

\subsection{Simulation Study \#2}\label{sec:sim2}
We simulated 100,000 teams by using rejection sampling: if the capability coefficients are within the interval [0.20, 0.32] for the squares cookie and [0.63, 1.00] for the letters cookie and meets the relationship between the capability and preference criteria for the specified team type then we keep the team. We focused on capability coefficients that are within these two intervals based on the results of the first user study (Section~\ref{sec:pilot_user1}). For each team, we solved for the optimal policies. The reward function for the efficient policy is defined as in Equation~\ref{eq:eff_reward_signal}, and the reward function for the fair-equity policy is defined as in Equation~\ref{eq:eff_fair_reward_signal}. 

\subsubsection{Results: Mixed Team Type}\label{sec:sim2_results_mixed_teams}
In this team type, $H0$ represents the person who have the highest capability in the team and their capabilities and task preferences are \textit{negatively} correlated. $H1$ represents the person who does \textit{not} have the highest capability in the team and their capabilities and task preferences are \textit{positively} correlated. Note that both humans are best at the same type of task which is making the letters cookie in the bakery scenario.   

\begin{figure}[!thb]
\centering
\includegraphics[width=0.8\linewidth]{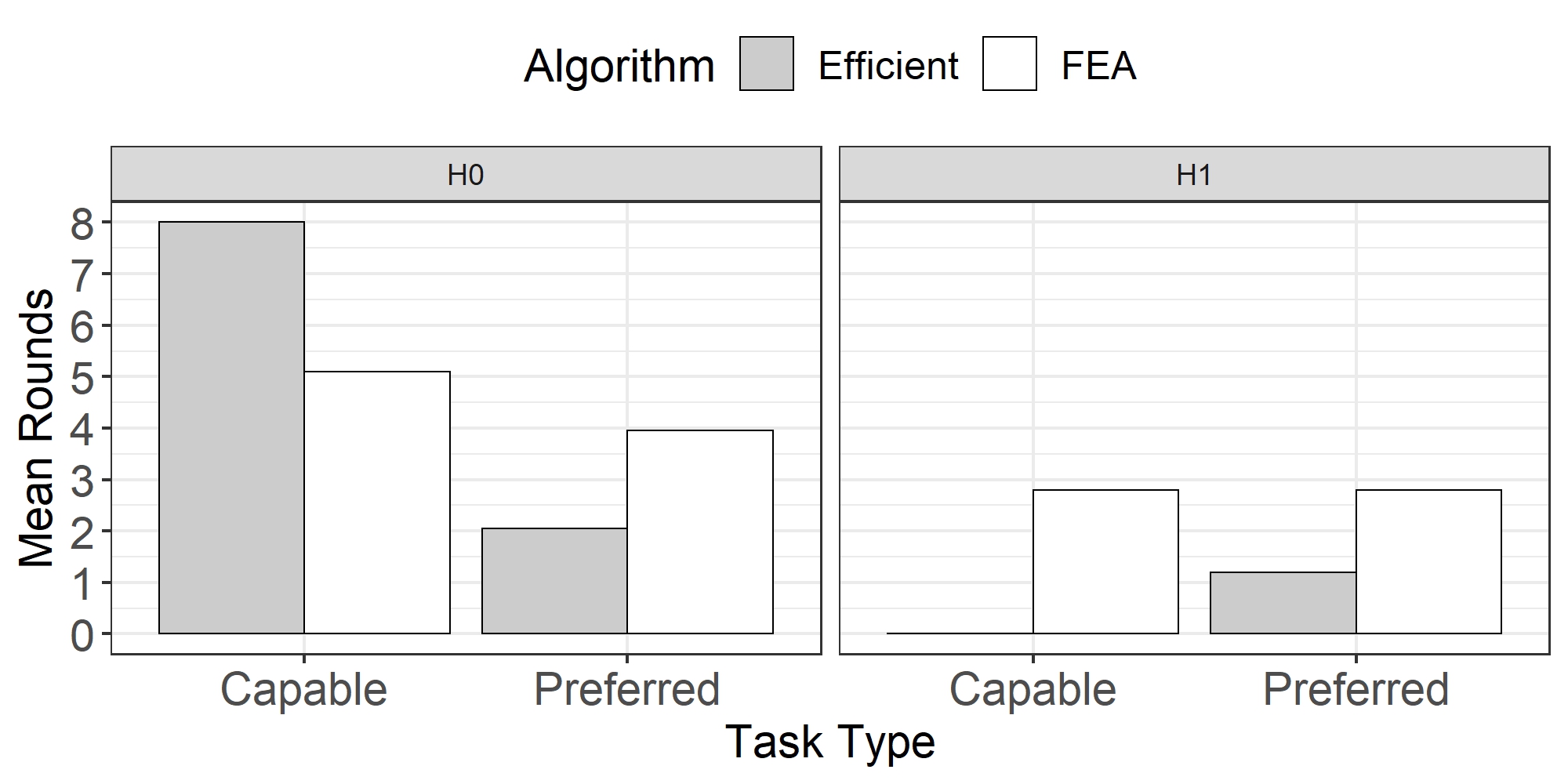}
\caption{Simulation study \#2 results for the Mixed team type.}
\label{fig:sim2_mixed_results}
\end{figure}

Figure~\ref{fig:sim2_mixed_results} shows the average number of rounds that each team member is assigned each task type. The results show that the Efficient algorithm assigns the person with the highest capability ($H_0$) to work on the task type that they are most capable at 100\% of the time. Since this person's capability is negatively correlated with their preferences, they end up getting to work on tasks they prefer about $25$\% of the time on average. On the other hand, FEA gives $H_0$ $50\%$ more time on average compared to the Efficient algorithm to work on their most preferred tasks. 

For the other person ($H_1$), the Efficient algorithm does not give them any chance to work on their most capable tasks and an average of $18\%$ of the time working on their most preferred tasks. On the other hand, FEA gives them about $34\%$ of the time to work on their most capable tasks and also about $34\%$ of the time to work on their most preferred task. Overall, FEA gives $H_1$ more opportunities to work on their most capable and preferred tasks.

\subsubsection{Results: Twins Team Type}\label{sec:sim2_results_twins_team}

As expected, the results show that both algorithms perform similarly; each person gets to work on their most capable tasks about $50\%$ of time and their most preferred tasks the rest of the time.

\subsubsection{Summary: FEA is Most Beneficial in the Mixed Team Type}
The simulation results reveal that FEA is the most beneficial to teams that belong to the Mixed team type. For the most capable team member in the Mixed team type, the Efficient algorithm gives them the most opportunities to work on their most capable tasks and the least opportunities to work on their preferred tasks compared to the FEA. For the other team member who is \textit{not} the most capable in the team, the Efficient algorithm does not give them any opportunities to work on their most capable tasks. In general, the difference between the average number of rounds that each team member gets to work on their most capable tasks and most preferred tasks is higher in the conditions with the Efficient algorithm than in the conditions with FEA. For the Twins team type, both algorithms allocate the same number of rounds for each team member to work on their most preferred and capable tasks.

%%%%%%%%%%%%%%%
\subsection{User Study \#2}
For User Study \#2, we evaluated FEA against the Efficient algorithm in the Mixed team type where the participant is the most capable in the team. We used the same study design as described in Section~\ref{sec:user_studies_overview} with modifications to the instructions and post-study questionnaire based on the lessons learned regarding transparency of the algorithms (Section~\ref{sec:implement_transparency}). A total of seven participants (3 female, 4 males, age: $M = 24.00$), participated in the study. Due to time constraints, we administered the modified post-study questionnaire to the last four participants.

%%%%%%%%%%%%%%%%%%%%
\subsubsection{Results}
Since this study has a small sample size ($n = 7$), we do not perform statistical tests. Data from the practice session shows that participants' average capability coefficient is $0.24$ for the squares cookie and $0.85$ for the letters cookie. Participants' average preference coefficient is $0.90$ for the squares cookie and $0.07$ for the letters cookie. We were able to better select the agent teammate compared to User Study \#1. The average $L_1$ value is $0.02$. 

\textbf{Task Performance:}
In general, it seems that teams in the Efficient condition ($M = 16.30$) completed more cookies than teams in the FEA condition ($M = 14.45$). Three out of the seven team ($43\%$) seem to perform similar in both conditions. This suggests that FEA can balance fairness and efficiency.

\textbf{Allocations:}
Also, participants seem to receive on average more rounds to work on their most capable tasks in the Efficient condition ($M = 8.00$) compared to the FEA condition ($M = 4.43$). The opposite is true for the agent teammates (Efficient: $M = 0.00$; FEA: $M = 3.57$). When interacting with the Efficient robot, both participants ($M = 0.00$) and agent teammates ($M = 2.29$) received less rounds to work on their most preferred tasks compared to when interacting with the FEA robot (participants: $M = 3.57$; agent teammates: $M = 3.00$). The patterns of these results are consistent with the Simulation Study \#2 results (Section~\ref{sec:sim2_results_mixed_teams}).

\begin{figure}[!thb]
\centering
\includegraphics[width=0.8\linewidth]{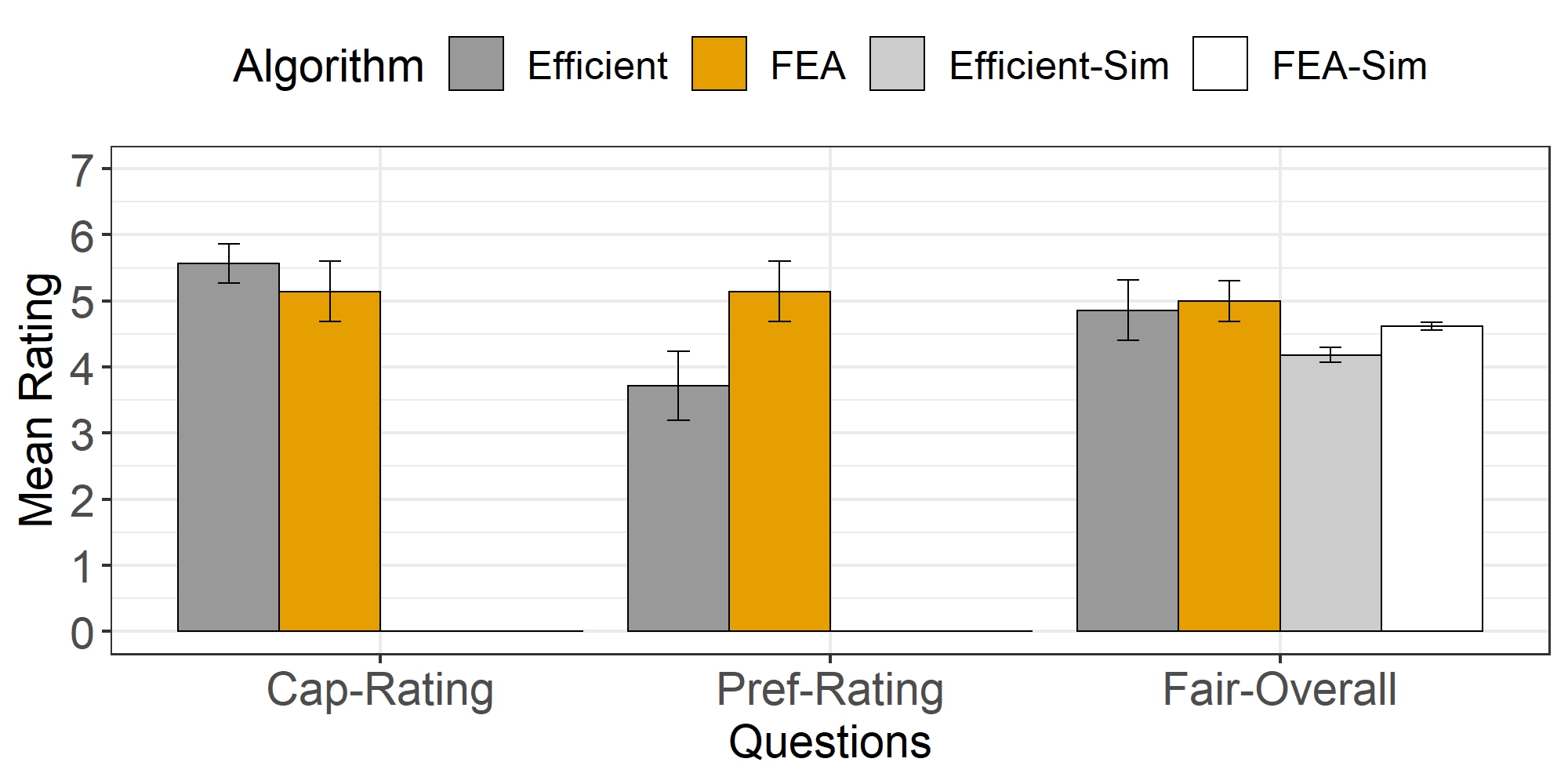}
\caption{Predicted fairness (scaled to rating between 1 and 7) from Simulation Study \#2 and participants' ratings from User Study \#2: Higher ratings mean more fair. Please see Table\ref{tab:pilot1_questionnaire1} for the specific questions. Cap is the abbreviation of capability. Pref is the abbreviation of preference.}
\label{fig:user2_ratings}
\end{figure}

\textbf{Perceived Fairness:}
Figure~\ref{fig:user2_ratings} show the results for perceived fairness and predicted fairness from Simulation Study \#2. The predicted fairness is the fair reward scaled to 1 to 7. 

\textbf{Fairness Based on Capabilities:} In both conditions, participants tend to highly agree that fairness based on capabilities are good fairness metrics (Efficient: $M = 6.00$; FEA: $M = 6.29$). They seem to also perceive both conditions to be fair in accounting for capabilities (Efficient: $M = 5.57$; FEA: $M = 5.14$).

\textbf{Fairness Based on Preferences:} Participants seem to agree about considering fairness with respect to preferences (Efficient: $M = 5.71$; FEA: $M = 6.00$). When asked to assess fairness based on preferences, they seem to perceive the FEA robot ($M = 5.14$) to be fairer than the Efficient one ($M = 3.71$). 

\textbf{Overall Perceived Fairness:} In terms of overall fairness, participants seem to feel that both robots were fair (Efficient: $M = 4.86$; FEA: $M = 5.00$).

\textbf{Generalization}
When asked to select which robot did a better job considering the team's preferences and cookie decoration speeds in its assignments, two out of the four participants vote for the Efficient robot and the other two vote for the FEA robot. The next time the tasks were to be completed, two of the four participants would want to work with the Efficient robot and the other two would like to to work with the FEA robot.

%%%%%%%%%%%%%%%
\subsubsection{Summary of FEA}
Teams working with the Efficient robot seem to complete more cookies than those working with the FEA robot. That is because the Efficient robot gives participants more opportunities to work on their most capable tasks and less opportunities to work on their most preferred tasks. However, participants tend to feel that both algorithms are fair with respect to capabilities. The FEA robot seems to give participants both opportunities to work on their most preferred and capable tasks. Participants seem to feel that the FEA robot is fairer in terms of taking into account preferences compared to the Efficient robot which indicates an improvement from the Fair algorithm in User Study \#1. Another improvement from the Fair algorithm is that participants seem to perceive FEA to be overall fair.

%%%%%%%%%%%%%%%%%%%%%%%%%%%%%%%%%%%%%%%%%%%%%%%%%%%%%%%%%%%%%%%%%%%%%%%%
\section{Discussion}
We extend an existing fairness metric~\cite{chang2021unfair} and use an iterative approach consisting of simulation and user studies to design an algorithm that balances fairness and efficiency in task allocations within human-agent teams. Based on our insights, we share design implications, open research questions, and future work. 

\subsection{Fairness}
We expand the equality of capability metric, $F_c$ from existing research to account for granular skill levels and propose equality of preference. Our work highlights a limitation to the equality of capability metric. Even though our simulation study results show a difference in fair rewards between the baseline algorithm and the Fair algorithm, our participants did not seem to perceive this difference. For example, our participants seem to think that both algorithms perform similarly to each other in terms of accounting for fairness based on preferences. They also seem to feel that both algorithms were unfair.  

In particular, participants who are the most capable in the team seem to experience unfairness from task allocations that equalize capabilities  across the team ($F_c$) which is a contrast to prior findings~\cite{chang2021unfair}. This may be because of the difference in POV. Prior work's assessments were from a first-person POV and our work is from a third-person POV. Moreover, our insights suggest that our participants may not have performed social perspective-taking, i.e., taking on another person's POV in their assessment of fairness. For instance, they may not have taken on the POV of an observer such as their supervisor. This observation contradicts the social science literature that people's judgment of fairness is independent of their POV~\cite{brugman2016fairness, galinsky2000perspective}.

Our participants who are the most capable in the team desire tasks to be distributed in a way that is proportional to their capabilities which we capture in our fair-equity metric. This insight aligns with the general trend in literature that team members want to be given more opportunities to contribute to the teamwork~\cite{jung2020robot, claure2020multi}. An open research question is how well fair-equity aligns with the perspective of the team member who is \textit{not} the most capable in the team.

Based on our insights, we recommend that designers and developers can leverage a stakeholder-centered approach~\cite{forlizzi2018moving} from HCI to design for the appropriate fairness metrics and optimization of fairness by considering the various stakeholders or POVs in the team. An interesting research question is how an agent can reason about fairness when there are conflicts among different POVs. 

\subsection{Fairness and Efficiency}
From this limitation of $F_c$, we implement five lessons learned in our next phase of simulation and user studies that includes developing the fair-equity metric and FEA. We saw in our simulation results that FEA is the most beneficial in the Mixed team type. Our follow-up user study shows promising results that participants seem to feel that FEA was fairer in terms of preferences and some seem to prefer to work with the FEA robot in the future. Some teams seem to be equally efficient in both conditions, which suggests that FEA can maintain efficiency while treating teammates fairly. This insight is different from prior research showing that there is generally trade-offs when accounting for fairness~\cite{dutta2020there, kim2020fact, menon2018cost,zhao2019inherent}. In addition, we advance prior work by providing a deeper understanding of fairness in relation to team members' characteristics and point of views.

Future work includes evaluating FEA within different team types in a set of user studies with more participants. Moreover, future research can explore generalizing the fairness metrics to $n$-team members. This is a challenge as people assess fairness by comparing their outcomes against others who are in a similar situation~\cite{runciman1966relative, bicchieri1999local_fairness}. ``Others'' could be defined as individuals such as team members within the same team. ``Others'' could also be defined as groups such as other teams.

%%%%%%%%%%%%%%%%%%%%%%%%%%%%%%%%%%%%%%%%%%%%%%%%%%%%%%%%%%%%%%%%%%%%%%%%
\section{Conclusion}
We used an iterative design approach to incorporate fairness and efficiency in our task allocation algorithm for effective human-agent teaming. We discovered that the existing equality of capability metric may be dependent on an individual's POV. This metric seems to align with perceived fairness from a third-person POV but not from a first-person POV. This led us to propose the fair-equity metric and incorporate it in our algorithm (FEA). Our insights show promise that FEA will be perceived to be both fair and efficient within different team types. We advance prior work by providing a deeper understanding of fairness in relation to team members' characteristics and point of views.

\begin{acks}
We thank student research assistants, Ty Brinker, Seema Kulkarni, and Hue Chang, for their work on the user study GUI.
\end{acks}

\bibliographystyle{ACM-Reference-Format}
\bibliography{refs}


\end{document}